\documentclass[fleqn,usenatbib]{mnras}

\usepackage{newtxtext,newtxmath}

\usepackage[T1]{fontenc}

\DeclareRobustCommand{\VAN}[3]{#2}
\let\VANthebibliography\thebibliography
\def\thebibliography{\DeclareRobustCommand{\VAN}[3]{##3}\VANthebibliography}

\usepackage{graphicx}	
\usepackage{amsmath}	
\usepackage{multirow,bigdelim}

\usepackage{xcolor}

\newcommand\cm{cm$^{-1}$}
\newcommand\km{km mol$^{-1}$}
\newcommand\um{$\mu$m}

\newcommand{\alo}{AlH$_3$OH$_2$}
\newcommand{\sio}{SiH$_3$OH}
\renewcommand{\sin}{SiH$_3$NH$_2$}

\title[AlH$_3$OH$_2$, SiH$_3$OH, and SiH$_3$NH$_2$]{Vibrational and
  Rotational Spectral Data for Possible Interstellar Detection of
  AlH$_3$OH$_2$, SiH$_3$OH, and SiH$_3$NH$_2$}

\author[A. G. Watrous et al.]{
  A. G. Watrous,$^{1}$
  B. R. Westbrook,$^{1}$
  M. C. Davis$^{1}$
  and Ryan C. Fortenberry$^{1}$\thanks{E-mail: r410@olemiss.edu}
  \\
  $^{1}$Department of Chemistry \& Biochemistry, University of
  Mississippi, University, MS 38677-1848, US }

\date{Accepted XXX. Received YYY; in original form ZZZ}

\pubyear{2021}

\begin{document}
\label{firstpage}
\pagerange{\pageref{firstpage}--\pageref{lastpage}}
\maketitle

\begin{abstract}

  This work provides the first full set of vibrational and rotational
  spectral data needed to aid in the detection of AlH$_3$OH$_2$,
  SiH$_3$OH, and SiH$_3$NH$_2$ in astrophysical or simulated
  laboratory environments through the use of quantum chemical
  computations at the CCSD(T)-F12b level of theory employing quartic
  force fields for the three molecules of interest. Previous work has
  shown that SiH$_3$OH and SiH$_3$NH$_2$ contain some of the strongest
  bonds of the most abundant elements in space. \alo\ also contains
  highly abundant atoms and represents an intermediate along the
  reaction pathway from H$_2$O and AlH$_3$ to AlH$_2$OH. All three of
  these molecules are also polar with AlH$_3$OH$_2$ having the largest
  dipole of 4.58 D and the other two having dipole moments in the
  1.10-1.30 D range, large enough to allow for the detection of these
  molecules in space through rotational spectroscopy. The molecules
  also have substantial infrared intensities with many of the
  frequencies being over 90 km mol$^{-1}$ and falling within the
  currently uncertain 12-17 \um\ region of the spectrum. The most
  intense frequency for AlH$_3$OH$_2$ is $\nu_9$ which has an
  intensity of 412 km mol$^{-1}$ at 777.0 cm$^{-1}$ (12.87
  \um). SiH$_3$OH has an intensity of 183 km mol$^{-1}$ at 1007.8
  cm$^{-1}$ (9.92 \um) for $\nu_5$, and SiH$_3$NH$_2$ has an intensity
  of 215 km mol$^{-1}$ at 1000.0 cm$^{-1}$ (10.00 \um) for $\nu_7$.

\end{abstract}




\section{Introduction}

Extraterrestrial Al-O bonds have been found in the form of AlOH and
AlO in evolved stars and in the atmosphere of the exoplanet WASP-43b
\citep{Tenenbaum09, Tenenbaum10, Takigawa17, Chubb20}. Additionally,
silicon monoxide (SiO) has been found in Sagittarius B2
\citep{Wilson71}, but the astronomical literature has not reported any
other small molecule detections containing an Si-O bond
\citep{McGuire18}. However, crystalline compounds containing silicon
and oxygen, such as enstatite (MgSiO$_3$), have been spectroscopically
observed toward NGC 6302 \citep{Molster01, Molster02} with most of the
spectral data falling in the range of 17-50 $\mu$m. In contrast, the
2.4-12 $\mu$m range of the spectrum is mostly attributed to carbon
molecules such as polycyclic aromatic hydrocarbons (PAHs)
\citep{Molster01}. Despite the typical attribution of this region to
carbonaceous PAHs, there are still many unknowns in the observed
spectra that may be filled by other Si containing molecules. In
particular, spectral data has many peaks from 12-17 $\mu$m that
correspond to unknown molecules, but this ``uncertain part of the
spectra'' \citep{Molster01} is in the range of many vibrational
features for known Si structures \citep{Molster01, Molster02}. Hence,
other Si containing molecules may yet be observed in that
region. Additionally, oxygen rich stars do not have nearly as strong
of PAH features allowing for Si containing molecules with strong
vibrational features to be detected there \citep{Molster02}.

Past work on inorganic oxides and related hydrides, including (MO)$_2$
with M = Al and Si, have been found to have anharmonic frequencies
below 20 $\mu$m (500 \cm) \citep{Westbrook20}. Most of the M-O bond
frequencies fall within the range of the unclear area of IR spectra,
and the metal oxides have intensities of over 50 km mol$^{-1}$ for
about half of the frequencies. Due to the number of frequencies in a
small range within this region, the intensities of the frequencies
need to be relatively strong to stand out from the forest of
lines. Additionally, the metal oxides and related hydrides previously
mentioned have similar features to large, crystalline forms of Si
containing molecules such as enstatite and forsterite (Mg$_2$SiO$_4$),
both of which have been detected in space \citep{Molster01,
  Molster02}. Enstatite and forsterite monomers have also been
reported to exhibit infrared intensities over 200 km mol $^{-1}$ which
is a large enough intensity for the frequencies of these molecules to
stand out from the spectral features of other molecules
\citep{Valencia20}. Consequently, this is a promising spectral region
in which to look for other silicon containing molecules.

Previous work has also shown that oxygen and nitrogen form bonds with
aluminum and silicon that are stronger than nearly any others in the
first three rows of the periodic table \citep{Doerksen20}. Of these
aforementioned strong bonds, aluminum and oxygen form the strongest
bond followed by silicon and oxygen. As such, AlH$_2$OH and
AlH$_2$NH$_2$ have been previously studied computationally and have
shown promise for interstellar detection due to their large dipole
moments and intense vibrational frequencies
\citep{Watrous21}. Additionally, a previously computed exothermic
pathway has shown that H$_2$O reacts with a simple aluminum hydride
(AlH$_3$) molecule to produce AlH$_2$OH which could easily form in
atmospheres of massive stars. This is due to the abundance of water
\citep{McGuire18} and the perceived presence of AlH$_3$ which is
likely unobserved because of its lack of a permanent dipole moment
\citep{Swinnen09}. AlH$_3$OH$_2$ has been previously computed to be an
intermediate formed barrierlessly along this reaction pathway and has
an energy lower than the starting reactants \citep{Swinnen09}. For
AlH$_3$OH$_2$ to persist on its own instead of serving as a
short-lived intermediate, it will have to dissipate the excess energy
from its formation by emission in the infrared. Such emission will
make the infrared spectral data provided herein invaluable for a
potential astronomical detection.  Regardless, if this molecule is
found in space, then AlH$_3$ and H$_2$O would also likely be present
in the gas phase, allowing the detection of AlH$_3$OH$_2$ to provide
evidence for the presence of the otherwise difficult-to-observe
AlH$_3$ molecule. SiH$_3$OH and SiH$_3$NH$_2$ are related molecules of
interest given the chemical similarity of Si and Al and the strength
of bonds between Si and N or O. Not only does silicon form strong
bonds with oxygen \citep{Doerksen20}, but it also is the third most
abundant element in the earth's mantle after oxygen and magnesium,
making it a potential component in the formation of planets from dust
grains.

To that end, this work utilizes quantum chemical methods to compute
optimized structures, energies, and spectroscopic data for
AlH$_3$OH$_2$, SiH$_3$OH, and SiH$_3$NH$_2$. The latter two containing
Si have relatively strong bonds of at least $-$91.10 kcal mol$^{-1}$
\citep{Doerksen20}, and all three contain highly abundant atoms
\citep{Savage96}. The Al-O bond in \alo\ is dative in nature but still
has an energy of more than 17 kcal mol$^{-1}$ \citep{Swinnen09} making
it usable with the present methods. None of the three molecules has
yet been detected in space, and neither \alo\ nor \sio\ have been
examined experimentally via gas-phase infrared spectroscopy in the
laboratory. On the other hand, \sin\ has been previously observed in
the infrared \citep{Beach92}, offering some gas-phase experimental
data to benchmark the present computational work. However, there are
still holes in the spectral interpretation that remain to be
elucidated.

Theoretical spectral data can be provided at a reasonable balance
between accuracy and computational cost by quartic force fields
(QFFs), which are a fourth-order Taylor series expansions of the
internuclear potential term of the Watson Hamiltonian
\citep{Fortenberry19QFF}. QFFs based on coupled cluster theory at the
singles, doubles, and perturbative triples level within the explicitly
correlated F12b construction [CCSD(T)-F12b] and with the corresponding
cc-pVTZ-F12 basis set \citep{Rag89O3, Shavitt09, ccreview, Peterson08,
  Yousaf08} typically produce fundamental vibrational frequencies
within 5 to 7 cm$^{-1}$ of gas-phase experiments \citep{Agbaglo19b,
  Agbaglo19c}. This level of accuracy is sufficient enough to lend
theoretical support to NASA missions such as the \textit{Stratospheric
  Observatory for Infrared Astronomy} (SOFIA) or the upcoming
\textit{James Webb Space Telescope} (JWST).

\section{Computational Details}

The geometry optimizations, harmonic frequencies, single point
energies, and dipole moments are computed using the Molpro 2015.1
\citep{Molpro15} and Molpro 2020.1 software packages \citep{MOLPRO} at
the CCSD(T)-F12b/cc-pVTZ-F12 \citep{Adler07, Peterson08, Yousaf08,
  Knizia09} level of theory. This will be referred to as F12-TZ. The
MP2/aug-cc-pVTZ double-harmonic intensities are computed using the
Gaussian16 program \citep{MP2, g16, aug-cc-pVXZ}. Each molecular
geometry is first optimized at the F12-TZ level of theory. This
geometry is then displaced by 0.005 \AA{} or 0.005 radians along the
symmetry-internal coordinates to form the QFFs.

\begin{figure}
  \centering
  \includegraphics[trim={100 150 250 700},clip,width=0.5\textwidth]{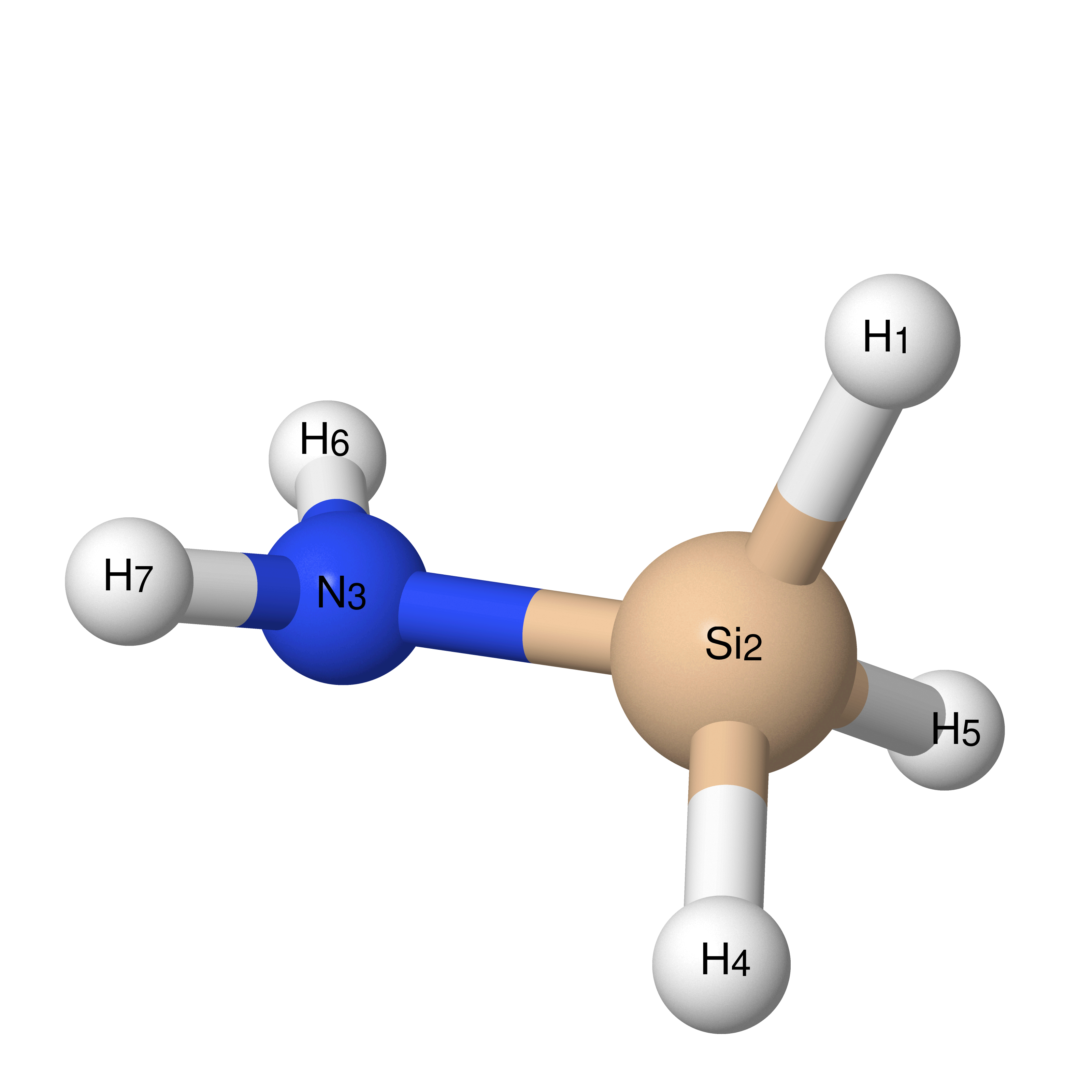}
  \caption{Visual depiction of the SiH$_3$NH$_2$ structure}
  \label{fig:sin}
\end{figure}

The QFF for SiH$_3$NH$_2$ is composed of 19585 points and is defined
by the coordinate system below, with atom labels corresponding to
Figure \ref{fig:sin}.

{
  \allowdisplaybreaks
  \begin{align}
    S_{ 1}(a') &= &r(\text{H}_1 - \text{Si}_2)\\
    S_{ 2}(a') &= &r(\text{Si}_2 - \text{N}_3)\\
    S_{ 3}(a') &= &\frac{1}{\sqrt{2}}[r(\text{Si}_2 - \text{H}_4) + r(\text{Si}_2 - \text{H}_5)]\\
    S_{ 4}(a') &= &\frac{1}{\sqrt{2}}[r(\text{N}_3 - \text{H}_6) + r(\text{N}_3 - \text{H}_7)]\\
    S_{ 5}(a') &= &\angle(\text{H}_1 - \text{Si}_2 - \text{N}_3)\\
    S_{ 6}(a') &= &\frac{1}{\sqrt{2}}[\angle(\text{H}_5 - \text{Si}_2 - \text{N}_3) + \angle(\text{H}_4 - \text{Si}_2 - \text{N}_3)]\\
    S_{ 7}(a') &= &\frac{1}{\sqrt{2}}[\angle(\text{H}_6 - \text{N}_3 - \text{Si}_2) + \angle(\text{H}_7 - \text{N}_3 - \text{Si}_2)]\\
    S_{ 8}(a') &= &\frac{1}{\sqrt{2}}[\tau(\text{H}_1 - \text{Si}_2 - \text{N}_3 - \text{H}_6) - \tau(\text{H}_1 - \text{Si}_2 - \text{N}_3 - \text{H}_7)]\\
    S_{ 9}(a') &= &\frac{1}{\sqrt{2}}[\tau(\text{H}_5 - \text{Si}_2 - \text{N}_3 - \text{H}_6) - \tau(\text{H}_4 - \text{Si}_2 - \text{N}_3 - \text{H}_7)]\\
    S_{10}(a'') &= &\frac{1}{\sqrt{2}}[r(\text{Si}_2 - \text{H}_4) - r(\text{Si}_2 - \text{H}_5)]\\
    S_{11}(a'') &= &\frac{1}{\sqrt{2}}[r(\text{N}_3 - \text{H}_6) - r(\text{N}_3 - \text{H}_7)]\\
    S_{12}(a'') &= &\frac{1}{\sqrt{2}}[\angle(\text{H}_5 - \text{Si}_2 - \text{N}_3) - \angle(\text{H}_4 - \text{Si}_2 - \text{N}_3)]\\
    S_{13}(a'') &= &\frac{1}{\sqrt{2}}[\angle(\text{H}_6 - \text{N}_3 - \text{Si}_2) - \angle(\text{H}_7 - \text{N}_3 - \text{Si}_2)]\\
    S_{14}(a'') &= &\frac{1}{\sqrt{2}}[\tau(\text{H}_1 - \text{Si}_2 - \text{N}_3 - \text{H}_6) + \tau(\text{H}_1 - \text{Si}_2 - \text{N}_3 - \text{H}_7)]\\
    S_{15}(a'') &= &\frac{1}{\sqrt{2}}[\tau(\text{H}_5 - \text{Si}_2 - \text{N}_3 - \text{H}_6) + \tau(\text{H}_4 - \text{Si}_2 - \text{N}_3 - \text{H}_7)]
  \end{align}
}

\noindent
The coordinate system for AlH$_3$OH$_2$ is the same as SiH$_3$NH$_2$
above with Al in the place of Si and O in the place of N.

\begin{figure}
  \centering
  \includegraphics[trim={600 400 100 400},clip,width=0.5\textwidth]{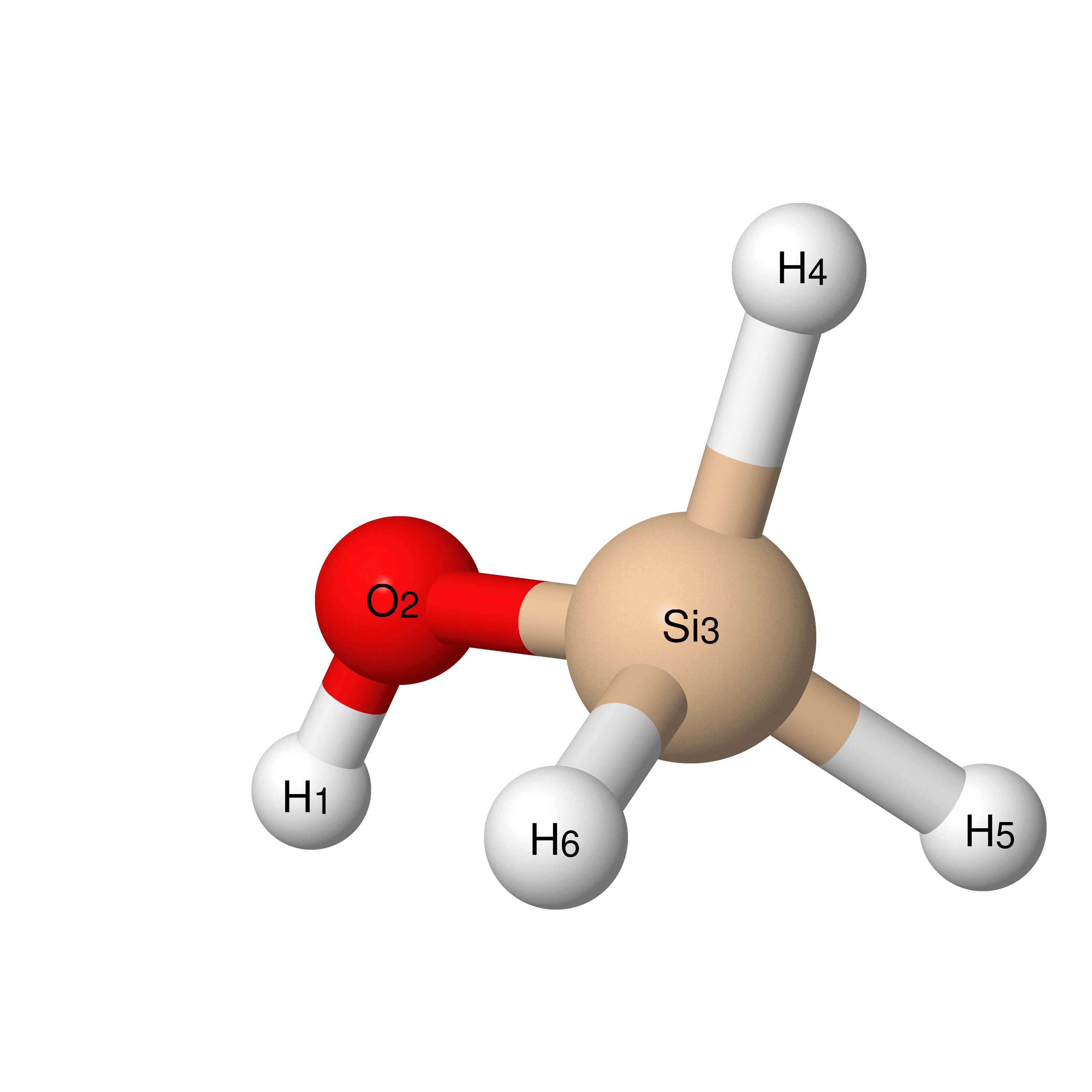}
  \caption{SiH$_3$OH}
  \label{fig:sio}
\end{figure}

The coordinate system for SiH$_3$OH requires 8481 points and is
defined in the following with atom labels corresponding to Figure \ref{fig:sio}.

\begin{align}
  S_{ 1}(a') &= &r(\text{H}_1 - \text{O}_2)\\
  S_{ 2} (a')&= &r(\text{O}_2 - \text{Si}_3)\\
  S_{ 3}(a') &= &r(\text{Si}_3 - \text{H}_4)\\
  S_{ 4}(a') &= &\frac{1}{\sqrt{2}}[r(\text{Si}_3 - \text{H}_5) + r(\text{Si}_3 - \text{H}_6)]\\
  S_{ 5} (a') &= &\angle(\text{H}_1 - \text{O}_2 - \text{Si}_3)\\
  S_{ 6}(a') &= &\angle(\text{O}_2 - \text{Si}_3 - \text{H}_4)\\
  S_{ 7}(a') &= &\frac{1}{\sqrt{2}}[\angle(\text{O}_2 - \text{Si}_3 - \text{H}_5) + \angle(\text{O}_2 - \text{Si}_3 - \text{H}_6)]\\
  S_{ 8}(a') &= &\frac{1}{\sqrt{2}}[\tau(\text{H}_1 - \text{O}_2 - \text{Si}_3 - \text{H}_5) - \tau(\text{H}_1 - \text{O}_2 - \text{Si}_3 - \text{H}_6)]\\
  S_{ 9}(a'') &= &\frac{1}{\sqrt{2}}[r(\text{Si}_3 - \text{H}_5) - r(\text{Si}_3 - \text{H}_6)]\\
  S_{10}(a'') &= &\tau(\text{H}_1 - \text{O}_2 - \text{Si}_3 - \text{H}_4)\\
  S_{11}(a'') &= &\frac{1}{\sqrt{2}}[\angle(\text{O}_2 - \text{Si}_3 - \text{H}_5) - \angle(\text{O}_2 - \text{Si}_3 - \text{H}_6)]\\
  S_{12}(a'') &= &\frac{1}{\sqrt{2}}[\tau(\text{H}_1 - \text{O}_2 - \text{Si}_3 - \text{H}_5) + \tau(\text{H}_1 - \text{O}_2 - \text{Si}_3 - \text{H}_6)]
\end{align}

Following the computation of single point energies at each of these
displaced geometries, the QFFs are generated by a least-squares
fitting procedure where the sum of the squared residuals is on the
order of 10$^{-16}$ a.u.$^2$ or less for all molecules. This yields
the equilibrium geometry, and a refit of the potential energy surface
zeros the gradients and produces the new equilibrium geometry along
with the corresponding force constants. The resulting force constants
are transformed into Cartesian coordinates using the INTDER program
\citep{INTDER} which are then used in second-order rotational and
vibrational perturbation theory \citep{Mills72} within the SPECTRO
software package \citep{spectro} in order to calculate rovibrational
spectral data \citep{Watson77, Papousek82}. Type-1 and 2 Fermi
resonances, Fermi polyads \citep{Martin97}, Coriolis resonances, and
Darling-Dennison resonances are taken into account to increase the
accuracy of the rovibrational data \citep{Martin95, Martin97}. The
Fermi resonances for each molecule can be found in Table S2.

\section{Results and Discussion}

\subsection{Geometries and rotational constants}

\begin{table}
  \centering
  \caption{Rotational constants and dipole moment of \alo, \sin, and
    \sio}
  \begin{tabular}{llrrrrr}
Constant      &   Units &        \alo &       \sin &       \sio \\
\hline       
$A_0$         &     MHz &     56168.1 &    68818.9 &    77174.6 \\
$B_0$         &     MHz &      9274.6 &    12646.5 &    13775.4 \\
$C_0$         &     MHz &      9130.9 &    12224.7 &    13553.9 \\
\hline       
$\Delta_{J}$  &     kHz &      23.495 &     11.122 &     13.574 \\
$\Delta_{K}$  &     kHz &     165.768 &    359.365 &    463.425 \\
$\Delta_{JK}$ &     kHz &      93.777 &    124.706 &    177.170 \\
$\delta_{J}$  &      Hz &     474.202 &    467.870 &    255.285 \\
$\delta_{K}$  &     kHz &  $-$544.657 & $-$307.033 & $-$658.864 \\
$\Phi_{J}$    &     mHz &   $-$91.250 &   $-$2.035 &   $-$2.752 \\
$\Phi_{K}$    &      Hz &      66.334 &     32.968 &     83.512 \\
$\Phi_{JK}$   &      Hz &      24.203 &      6.555 &     23.787 \\
$\Phi_{KJ}$   &      Hz &   $-$84.242 &  $-$28.812 &  $-$87.728 \\
$\phi_{j}$    & $\mu$Hz & $-$1768.830 & $-$418.967 &    613.039 \\
$\phi_{jk}$   &      Hz &    $-$8.133 &   $-$1.303 &   $-$2.708 \\
$\phi_{k}$    &     kHz &       6.367 &      0.951 &      4.257 \\
\hline       
$\mu$         &       D &        4.58 &       1.21 &       1.29 \\
             
  \end{tabular}
  \label{tab:rot}
\end{table}

AlH$_3$OH$_2$ has the largest dipole moment out of the three molecules
studied at 4.58 D. This dipole moment is stronger than other, similar
aluminum and oxygen containing molecules such as AlH$_2$OH and AlOH
which have dipole moments of 1.22 and 1.11 D, respectively
\citep{Watrous21, Fortenberry20AlOH}. SiH$_3$OH has the second largest
dipole at 1.29 D which is much smaller than the dipole moment of
AlH$_3$OH$_2$. SiH$_3$NH$_2$ has a slightly smaller dipole moment
still than SiH$_3$OH at 1.21 D. All three of these molecules have
dipole moments large enough to be detectable in space through pure
rotational spectroscopy, but AlH$_3$OH$_2$ does have a significantly
larger dipole moment than the other two molecules. AlH$_3$OH$_2$,
SiH$_3$OH, and SiH$_3$NH$_2$ all show near-prolate character as
indicated by their $\kappa$ values of approximately $-0.99$. Given
this prolate nature, \sio\ and \sin\ could be expected to have
$B_{\text{eff}}$ rotational constants comparable to their diatomic
counterparts, SiO and SiN. However, in actuality the addition of the
hydrogen atoms causes the rotational constants nearly to halve
relative to the isolated diatomics \citep{Herzberg79} and renders such
a comparison moot. The principal rotational constants, quartic and
sextic distortion coefficients, and dipole moments are reported in
Table \ref{tab:rot}. The full set of equilibrium and first
vibrationally excited rotational constants is shown in Table S1.

\begin{table}
  \centering
  \caption{Vibrationally averaged and equilibrium geometrical
    parameters of AlH$_3$OH$_2$}
  \begin{tabular}{lrrr}
                                   & Units &       Value \\
\hline                            
R$_0$(H$_1$-Al$_2$)                 &  \AA{}  & 1.59197 \\
R$_0$(Al$_2$-O$_3$)                 &  \AA{}  & 2.04014 \\
R$_0$(Al$_2$-H$_4$)                 &  \AA{}  & 1.60154 \\
R$_0$(O$_3$-H$_6$)                  &  \AA{}  & 0.91884 \\
$\angle$$_0$(Al$_2$-H$_1$-O$_3$)    &    deg  & 101.834 \\
$\angle$$_0$(Al$_2$-H$_5$-O$_3$)    &    deg  & 96.206 \\
$\angle$$_0$(O$_3$-H$_6$-Al$_2$)    &    deg  & 116.385 \\
\hline                            
R$_e$(H$_1$-Al$_2$)                 &  \AA{}  & 1.58742 \\
R$_e$(Al$_2$-O$_3$)                 &  \AA{}  & 2.02504 \\
R$_e$(Al$_2$-H$_4$)                 &  \AA{}  & 1.59756 \\
R$_e$(O$_3$-H$_6$)                  &  \AA{}  & 0.96189 \\
$\angle_e$(Al$_2$-H$_1$-O$_3$)      &    deg  & 102.741 \\
$\angle_e$(Al$_2$-H$_5$-O$_3$)      &    deg  & 95.695 \\
$\angle_e$(O$_3$-H$_6$-Al$_2$)      &    deg  & 112.678 \\
  \end{tabular}
  \label{Geom_AlH3OH2}
\end{table}

As shown in Table \ref{Geom_AlH3OH2}, AlH$_3$OH$_2$ has a heavy atom
bond length of 2.03 \AA{}. This is longer than the Al-O bond in
AlH$_2$OH and AlOH which are 1.70 and 1.68 \AA{}, respectively
\citep{Watrous21, Fortenberry20AlOH}. In fact, this bond length is
even longer than the 1.67 \AA{} dative bond between NH$_3$ and BH$_3$
in ammonia borane \citep{Thorne83, Westbrook21}, suggesting that it
also is dative in nature. SiH$_3$OH has an Si-O bond length of 1.65
\AA{} which is shorter than the Al-O bond length of 1.70 \AA{} in
AlH$_2$OH and the Mg-O bond length of 1.77 \AA{} in HMgOH
\citep{Watrous21}. Despite the fact that Al-O has previously been
reported to have a stronger bond than Mg-O or Si-O, \citep{Doerksen20}
the slightly shorter bond length of Si-O compared to Al-O implies that
Si-O has a stronger bond in these molecules. However, this is a small
difference and the bond lengths of Si-O and Al-O are close to the same
value.  In SiH$_3$NH$_2$, the Si-N bond has a length of 1.74 \AA{}
which is smaller than the N-Mg bond length of 1.91 \AA{} in HMgNH$_2$
and slightly smaller than the Al-N bond length of 1.78 \AA{} in
AlH$_2$NH$_2$ \citep{Watrous21}. Just like with the aforementioned
Si-O bond, the Si-N bond is shorter than the previously studied Al-N
even though earlier work shows that the Al-N bond is stronger than the
Si-N bond \citep{Doerksen20}.

\begin{table}
  \centering
  \caption{Vibrationally averaged and equilibrium geometrical
    parameters of SiH$_3$OH }
  \begin{tabular}{lrrr}
                               & Units &       Value \\
\hline                        
R$_0$(H$_1$-O$_2$)              &  \AA{}  & 0.92024 \\
R$_0$(O$_2$-Si$_3$)             &  \AA{}  & 1.65312 \\
R$_0$(Si$_3$-H$_4$)             &  \AA{}  & 1.48371 \\
R$_0$(Si$_3$-H$_5$)             &  \AA{}  & 1.49070 \\
$\angle_0$(O$_2$-H$_1$-Si$_3$)  &    deg  & 119.969 \\
$\angle_0$(Si$_3$-O$_2$-H$_4$)  &    deg  & 106.489 \\
$\angle_0$(Si$_3$-O$_2$-H$_5$)  &    deg  & 111.282 \\
\hline
R$_e$(H$_1$-O$_2$)              &  \AA{}  & 0.95631 \\
R$_e$(O$_2$-Si$_3$)             &  \AA{}  & 1.64771 \\
R$_e$(Si$_3$-H$_4$)             &  \AA{}  & 1.47389 \\
R$_e$(Si$_3$-H$_5$)             &  \AA{}  & 1.48160 \\
$\angle_e$(O$_2$-H$_1$-Si$_3$)  &    deg  & 118.225 \\
$\angle_e$(Si$_3$-O$_2$-H$_4$)  &    deg  & 105.768 \\
$\angle_e$(Si$_3$-O$_2$-H$_5$)  &    deg  & 111.547 \\
  \end{tabular}
  \label{Geom_SiH3OH}
\end{table}

\begin{table}
  \caption{Vibrationally averaged and equilibrium geometrical
    parameters of SiH$_3$NH$_2$}
  \centering
  \begin{tabular}{lrrr}
                                 & Units &       Value \\
\hline                          
R$_0$(H$_1$-Si$_2$)               &  \AA{}  & 1.49434 \\
R$_0$(Si$_2$-N$_3$)               &  \AA{}  & 1.72495 \\
R$_0$(Si$_2$-H$_4$)               &  \AA{}  & 1.48713 \\
R$_0$(N$_3$-H$_6$)                &  \AA{}  & 0.98636 \\
$\angle_0$(Si$_2$-H$_1$-N$_3$)    &    deg  & 114.848 \\
$\angle_0$(Si$_2$-H$_5$-N$_3$)    &    deg  & 108.286 \\
$\angle_0$(N$_3$-H$_6$-Si$_2$)    &    deg  & 121.756 \\
\hline
R$_e$(H$_1$-Si$_2$)               &  \AA{}  & 1.48654 \\
R$_e$(Si$_2$-N$_3$)               &  \AA{}  & 1.72159 \\
R$_e$(Si$_2$-H$_4$)               &  \AA{}  & 1.47930 \\
R$_e$(N$_3$-H$_6$)                &  \AA{}  & 1.00706 \\
$\angle_e$(Si$_2$-H$_1$-N$_3$)    &    deg  & 115.666 \\
$\angle_e$(Si$_2$-H$_5$-N$_3$)    &    deg  & 107.725 \\
$\angle_e$(N$_3$-H$_6$-Si$_2$)    &    deg  & 119.860 \\
  \end{tabular}
  \label{Geom_SiH3NH2}
\end{table}

\subsection{Vibrational Frequencies}

\begin{table}
  \scriptsize
  \centering
  \caption{Vibrational Frequencies (in cm$^{-1}$) of AlH$_3$OH$_2$
    with MP2/aug-cc-pVTZ Intensities in Parentheses (in km
    mol$^{-1}$)}
  \begin{tabular}{llrr}
& Description & Harmonic & Anharmonic\\
\hline
$\nu_1$ $(a'')$ &1.001$S_{11}$& 3900.9 (142) & 3715.0\\
$\nu_2$ $(a')$ &1.001$S_{4}$& 3801.2 (62) & 3624.5\\
$\nu_3$ $(a')$ &0.952$S_{1}$+0.049$S_{3}$& 1920.4 (168) & 1854.5\\
$\nu_4$ $(a')$ &0.951$S_{3}$-0.049$S_{1}$ & 1881.9 (194) & 1819.3\\
$\nu_5$ $(a'')$ &1.001$S_{10}$& 1871.5 (367) & 1809.0\\
$\nu_6$ $(a')$ &0.622$S_{8}$+0.386$S_{7}$& 1649.2 (91) & 1607.4\\
$\nu_7$ $(a'')$ &0.555$S_{14}$-0.394$S_{15}$-0.053$S_{13}$& 796.5 (261) & 785.1\\
$\nu_8$ $(a')$ &0.475$S_{9}$-0.270$S_{6}$-0.159$S_{8}$-0.086$S_{5}$& 794.2 (297) & 778.9\\
$\nu_9$ $(a')$ &0.461$S_{6}$+0.349$S_{9}$+0.186$S_{5}$& 777.0 (412) & 775.3\\
$\nu_{10}$ $(a'')$ &0.548$S_{13}$+0.430$S_{12}$& 713.4 (5) & 664.0\\
$\nu_{11}$ $(a')$ &0.540$S_{5}$+0.186$S_{7}$-0.159$S_{6}$-0.092$S_{9}$& 523.4 (153) & 487.9\\
$\nu_{12}$ $(a')$ &0.430$S_{7}$-0.192$S_{5}$-0.176$S_{8}$+0.116$S_{6}$& 386.5 (148) & 319.7\\
$\nu_{13}$ $(a')$ &1.016$S_{2}$& 372.6 (23) & 350.6\\
$\nu_{14}$ $(a'')$ &0.542$S_{12}$-0.369$S_{13}$-0.093$S_{14}$& 366.0 (5) & 364.1\\
$\nu_{15}$ $(a'')$ &0.613$S_{15}$+0.328$S_{14}$& 94.7 (37) & 225.8\\
  \end{tabular}
  \label{tab:vib_alo}
\end{table}

Over half of the double-harmonic MP2/aug-cc-pVTZ intensities for
AlH$_3$OH$_2$, SiH$_3$OH, and Si$_3$NH$_2$ from Tables
\ref{tab:vib_alo}, \ref{tab:vib_sio}, and \ref{tab:vib_sin} are on the
same order or larger than the antisymmetric stretch in water, which
has an intensity of roughly 70 km mol$^{-1}$.  AlH$_3$OH$_2$, in
particular, has four anharmonic frequencies with intensities over 200
km mol$^{-1}$, $\nu_5$, $\nu_7$, $\nu_8$, and $\nu_9$ at 1809.0,
785.1, 778.9, and 775.3 cm$^{-1}$, respectively. These latter three
are all within the 12-17 \um\ region of the spectrum, making them
particularly useful for potential detection of \alo\ in this
window. Further, of \alo's five additional frequencies over 100 km
mol$^{-1}$, $\nu_{11}$ is also close to this window with a frequency
of 523.4 \cm\ or 19.11 \um.

The intense $\nu_5$ antisymmetric Al-H stretching frequency at 1809.0
cm$^{-1}$ corresponds to 5.53 $\mu$m which is in the range of the
infrared typically associated with PAHs \citep{Molster01}. This is
quite a bit lower in frequency relative to the Ar matrix experimental
value for the antisymmetric stretch of planar AlH$_3$ at 1882.7 \cm\
\citep{Kurth93}. Similarly, the observed value of the AlH$_3$ umbrella
motion at 697.6 cm$^{-1}$ is substantially lower than that of the
$\nu_{9}$ O-Al-H bend in AlH$_3$OH$_2$ at 775.3 cm$^{-1}$ due to the
water molecule's inhibition of this motion. In contrast, the
experimental frequency of the bend in AlH$_3$ at 783.5 cm$^{-1}$
compares very well to the $\nu_8$ H-O-Al-H torsion in AlH$_3$OH$_2$
which has a frequency of 778.9 cm$^{-1}$ implying a slight red-shift
of this motion upon complexation with water.

\begin{table}
  \scriptsize
  \centering
  \caption{Vibrational Frequencies (in cm$^{-1}$) of SiH$_3$OH with
    MP2/aug-cc-pVTZ Intensities in Parentheses (in km mol$^{-1}$)}
  \begin{tabular}{llrr}
& Description & Harmonic & Anharmonic\\
\hline
$\nu_1$ $(a')$ & 0.999$S_{1}$& 3919.8 (82) & 3742.9\\
$\nu_2$ $(a')$ &1.000$S_{3}$& 2286.9 (103) & 2202.0\\
$\nu_3$ $(a')$ &1.001$S_{4}$& 2245.5 (87) & 2167.4\\
$\nu_4$ $(a'')$ &1.001$S_{9}$& 2238.3 (162) & 2159.1\\
$\nu_5$ $(a')$ &0.722$S_{7}$+0.192$S_{6}$-0.047$S_{8}$& 1007.8 (183) & 990.6\\
$\nu_6$ $(a')$ &0.865$S_{8}$+0.065$S_{6}$-0.049$S_{5}$& 987.6 (113) & 966.4\\
$\nu_7$ $(a'')$ &0.665$S_{10}$-0.304$S_{12}$& 967.9 (79) & 952.3\\
$\nu_8$ $(a')$ &0.429$S_{2}$+0.341$S_{5}$+0.226$S_{6}$& 911.2 (70) & 889.9\\
$\nu_9$ $(a')$ &0.521$S_{2}$-0.314$S_{5}$-0.094$S_{6}$+0.069$S_{7}$& 850.0 (126) & 810.9\\
$\nu_{10}$ $(a'')$ &0.961$S_{11}$-0.046$S_{12}$& 729.5 (60) & 712.9\\
$\nu_{11}$ $(a')$ &0.422$S_{6}$-0.286$S_{5}$-0.214$S_{7}$-0.077$S_{8}$& 690.1 (58) & 683.3\\
$\nu_{12}$ $(a'')$ &0.651$S_{12}$+0.342$S_{10}$& 196.2 (93) & 181.1\\
  \end{tabular}
  \label{tab:vib_sio}
\end{table}

SiH$_3$OH similarly has five anharmonic frequencies with intensities
over 100 km mol$^{-1}$. Two of these, $\nu_2$ and $\nu_4$, correspond
to the symmetric and antisymmetric Si-H stretches. These frequencies
at 2202.0 and 2159.1 cm$^{-1}$ are equivalent to 4.54 and 4.63 $\mu$m
respectively and are, again, at the end of the infrared spectrum
typically associated with PAHs \citep{Molster01}. Another frequency
with a large intensity is $\nu_9$ with an intensity of 126 km
mol$^{-1}$ at 810.9 cm$^{-1}$ or 12.33 $\mu$m. This frequency is
mostly made of the Si-O stretch and is just within the 12-17 \um\
region of the spectrum, making it a prime target for detection of this
molecule. $\nu_{10}$ and $\nu_{11}$ also fall within this region at
712.9 and 683.3 \cm\ or 14.03 and 14.63 \um\, respectively, although
they have more modest intensities of 60 and 58 \km. Regardless, these
frequencies may also help to elucidate the presence of \sio\ in this
spectral window.

The SiH$_3$OH harmonic frequencies calculated in this work correspond
well to previous CCSD(T)/d-aug-cc-pVTZ results \citep{cccbdb}, with
the largest difference occurring in $\omega_5$ at just under 15
\cm. However, the average magnitude of the deviation is only 5.1
\cm. While there are no current experimental data for SiH$_3$OH,
isovalent CH$_3$OH is a well-known molecule in space \citep{Ball70,
  McGuire18Census} with comparable experimental data
\citep{Shimanouchi72}. As such, the anharmonic frequencies of CH$_3$OH
are also calculated at the F12-TZ level to serve as an additional
benchmark for the data generated for the other molecules in this
work. Of the 10 available gas-phase frequencies, the largest deviation
occurs in $\nu_{3}$ at a $-$54.0 \cm\ difference between the
experimental value of 2960 \cm\ and the F12-TZ result at 2906.0
\cm. However, the computed O-H stretch has a difference of only 0.8
cm$^{-1}$ from the experimental data, and the average magnitude of the
deviation is less than 10 \cm. This good overall agreement suggests
the performance of the same methodology should provide accurate
predictions of the fundamental frequencies of SiH$_3$OH, as well.

\begin{table}
  \scriptsize
  \centering
  \caption{Vibrational frequencies (in \cm) of \sin\ with
    MP2/aug-cc-pVTZ intensities (in km mol$^{-1}$) in parentheses}
  \begin{tabular}{llrrl}
                   &                               Description &     Harmonic &               Anharmonic & Experiment$^a$\\
                  
\hline            
$\nu_1$ $(a'')$    &                             1.001$S_{11}$ &  3677.8 (22) &                   3515.3 &          3547 \\
$\nu_2$ $(a')$     &                              1.000$S_{4}$ &  3584.8 (15) &                   3438.7 &          3445 \\
$\nu_3$ $(a'')$    &                             1.001$S_{10}$ & 2258.0 (141) & 2172.2\rdelim\}{3}{-5mm} &               \\
$\nu_4$ $(a')$     &                              0.977$S_{3}$ &  2256.1 (76) &                   2174.5 &     2172      \\
$\nu_5$ $(a')$     &                              0.978$S_{1}$ & 2208.9 (167) &                   2132.5 &               \\
$\nu_6$ $(a')$     &                 0.783$S_{7}$-0.218$S_{8}$ &  1598.0 (29) &                   1563.7 &          1564 \\
$\nu_7$ $(a')$     &    0.517$S_{6}$+0.387$S_{5}$+0.078$S_{9}$ & 1000.0 (215) &  984.7\rdelim\}{3}{-5mm} &          ~996 \\
$\nu_8$ $(a'')$    & 0.546$S_{14}$-0.334$S_{15}$-0.115$S_{12}$ &   997.5 (65) &                    976.0 &          ~983 \\
$\nu_9$ $(a')$     &                 0.903$S_{9}$-0.107$S_{6}$ &  944.4 (103) &                    932.4 &          ~970 \\
$\nu_{10}$ $(a'')$ &               0.623$S_{13}$-0.352$S_{12}$ &   921.1 (40) &                    885.8 &               \\
$\nu_{11}$ $(a')$  &                              0.938$S_{2}$ &   851.0 (49) &                    842.4 &          ~845 \\
$\nu_{12}$ $(a')$  &    0.612$S_{5}$-0.346$S_{6}$-0.072$S_{9}$ &   712.8 (63) &                    699.2 &      ~670$^b$ \\
$\nu_{13}$ $(a'')$ & 0.495$S_{12}$+0.368$S_{13}$+0.152$S_{14}$ &   629.9 (35) &                    608.7 &      ~670$^b$ \\
$\nu_{14}$ $(a')$  &    0.833$S_{8}$+0.197$S_{7}$+0.044$S_{9}$ &  393.4 (153) &                    244.2 &               \\
$\nu_{15}$ $(a'')$ &               0.683$S_{15}$+0.274$S_{14}$ &   177.5 (13) &                 242.8 \\
  \end{tabular}
  $^a$From Ref. \citenum{Beach92}\\
  $^b$Ref. \citenum{Beach92} points to both modes as potential
  identities for the peak at 670 \cm
  \label{tab:vib_sin}
\end{table}


Like SiH$_3$OH, SiH$_3$NH$_2$ has five anharmonic frequencies with
intensities over 100 km mol$^{-1}$. These are also the same types as
the five most intense frequencies in \alo: the three third-row atom
hydride stretches and two bends. The Si-H stretches fall just above
the corresponding Si-H stretches of SiH$_3$OH at 4.60 and 4.69 $\mu$m
or 2174.5 and 2132.5 \cm. Additionally, the $\nu_{14}$ torsional
fundamental at 244.2 \cm\ or 40.95 $\mu$m has an intensity of 153 km
mol$^{-1}$. This frequency is significantly farther out from the
infrared spectrum than the other frequency that has been attributed to
silicates between 2.4 and 12 \um\ \citep{Molster01}, offering a new
region for detecting such molecules at THz frequencies. The picture
within the 12-17 \um\ window is less vivid than for the other
molecules, however. Only $\nu_{12}$ and $\nu_{13}$ fall directly
within this region at 699.2 and 608.7 \cm\ or 14.30 and 16.43 \um, and
they have fairly low intensities of 63 and 35 \km. $\nu_{11}$ also
comes close to this window at 842.4 \cm\ or 11.87 \um, but it, too,
has a modest intensity of only 49 \km.  Still, all of these
frequencies may help to shed some light on this currently
sparsely-identified region of the spectrum.

The \sin\ vibrational frequencies reported in this paper compare well
to the available experimental data from \cite{Beach92}. In particular,
the computed $\nu_{3}$ antisymmetric Si-H stretch at 2172.2 \cm\ is
virtually identical to the gas-phase value of 2172 \cm.  However, the
fairly broad peak in the experimental spectrum could also be
attributed to $\nu_{4}$, the symmetric Si-H stretch, or even
$\nu_{5}$, the apical Si-H stretch, reported here to have very similar
frequencies of 2174.5 and 2132.5 \cm. The proximity of these three
modes would require a much more high-resolution experiment to separate
them, but our methods are giving unique identification for this
spectral region for this molecule. Similarly, $\nu_{7}$ and $\nu_{8}$
are assigned as shoulders on the more prominent $\nu_{9}$ peak around
970 \cm\ in the experimental spectrum. In the same vein, the very
broad peak centered around 670 \cm\ does not afford the opportunity to
separate the distinct peaks reported here at 699.2 and 608.7 \cm\ for
$\nu_{12}$ and $\nu_{13}$, respectively. In all of these cases, direct
attribution of the experimental spectrum is currently impossible, but
the F12-TZ values presented herein may help to identify the unique
spectral features of these modes in future high-resolution
experiments.

The related SiO \citep{Wilson71, Dickinson72} and SiN
\citep{Turner92a} molecules are well known in the interstellar medium
and potentially offer a point of comparison for the Si-O stretch in
\sio\ and the Si-N stretch in \sin. However, the Si-O stretch computed
herein for SiH$_3$OH (again, $\nu_9$) is substantially red-shifted
compared to the experimental value for SiO at 1229.6 cm$^{-1}$
\citep{Herzberg79}. Similarly, there is a large red-shift in the Si-N
stretch in \sin\ at 842.4 \cm\ compared to the experimental value for
isolated SiN at 1138.4 \cm\ \citep{Herzberg79}. This suggests that
even though the two pairs of molecules may be likely to occur in the
same regions, they should have little spectral overlap. In fact, the
closest \sio\ frequency to the SiO stretch is $\nu_5$ at 990.6 \cm,
and the closest \sin\ frequency to the SiN stretch is $\nu_{7}$ at
984.7 \cm, allowing \sio\ and \sin\ to be easily disentangled
spectroscopically from their related diatomic molecules, even if they
are found in the same region. This is especially important for \sio\
in the case that it has a similar reduction pathway as carbon monoxide
does to methanol.

Finally, $\nu_{15}$ for AlH$_3$OH$_2$ and SiH$_3$NH$_2$ both have
positive anharmonicities. In particular, $\nu_{15}$ of \alo\ shows
quite a large positive anharmonicity of 131.1 cm$^{-1}$ above the
harmonic frequency, yielding a fundamental frequency of 225.8
cm$^{-1}$. The positive anharmonicity for SiH$_3$NH$_2$ is not as
large with the anharmonic frequency at 242.8 cm$^{-1}$ only 65.3
cm$^{-1}$ above the harmonic frequency. While having such large
positive anharmonicities raises questions about the reliability of the
affected mode, it has been previously reported that near prolate
molecules show positive anharmonicities in the smaller frequencies
\citep{Wilhelm13, Fortenberry12HOCO+, Fortenberry12HOCS+,
  Huang13NNOH+, Watrous21}. These frequencies also have relatively
small intensities, suggesting that they would be unhelpful for
astronomical or laboratory detection anyway.

\section{Conclusions}

All three of the molecules examined in this work, AlH$_3$OH$_2$,
SiH$_3$OH, and SiH$_3$NH$_2$, have substantial dipole moments that
make them observable radioastronomically. In particular, AlH$_3$OH$_2$
has the largest dipole moment of the three at 4.58 D, while SiH$_3$OH
and Si$_3$NH$_2$ have smaller dipoles at 1.29 and 1.21 D
respectively. Each of these molecules also has at least five
fundamental frequencies with large intensities greater than 100 km
mol$^{-1}$, making the readily visible to infrared spectroscopic
investigation, such as that performed by JWST or by the ongoing SOFIA
mission. Nearly all of these intense modes are within the spectral
range of the high-resolution EXES instrument on SOFIA, making it a
particularly appealing avenue for taking advantage of the data
presented herein. Even more importantly, each of these molecules has
at least two frequencies within the uncertain 12-17 \um\ spectral
region, where identification may be particularly facile. For \sin\,
the frequencies in this region have intensities of only about 50 \km,
but \sio\ has one frequency just inside this window at 12.33 \um\ with
a substantial intensity of 126 \km. On the other hand, \alo\ has three
very intense frequencies in this region, with $\nu_{7}$, $\nu_{8}$,
and $\nu_{9}$ all having intensities above 261 \km, and $\nu_{9}$ in
particular having an enormous intensity of 412 \km. As such, \alo\ may
be particularly identifiable in this region.

Further, previous work has shown that AlH$_3$OH$_2$ is likely to be
formed from reactions of water and the possible interstellar molecule
of AlH$_3$ and may also provide insight into the formation of
inorganic oxides from simple metal hydrides and water
\citep{Swinnen09}. Hence, the spectral data reported here will help
with possible detection of this molecule through rotational and
infrared spectroscopy, especially given the very large intensities of
many of its fundamental frequencies and the occurrence of these
frequencies in a relatively unidentified spectral region. Such a
detection would help to shed some light on the abundance of the
rotationally-dark AlH$_3$.

SiH$_3$OH and SiH$_3$NH$_2$ are similar to Al and Mg compounds that
have been studied previously, and they both have highly intense
vibrational frequencies in their own rights. Namely, the Si-O stretch
in \sio\ at 850.0 \cm\ or 11.76 $\mu$m has an intensity of 126 km
mol$^{-1}$, and the Si-H stretches of \sin\ have intensities of 141,
76, and 167 \km\ at frequencies of 2258.0, 2256.1, and 2208.9 \cm\ or
4.429, 4.432, and 4.527 $\mu$m, respectively. Although these
frequencies do not fall in the 12-17 \um\ region like some of the
slightly less intense frequencies for these molecules, they are still
found in a region of the spectrum in which few molecules have
currently been identified and those that have been elucidated are
primarily carbonaceous molecules like PAHs. Hence, the present work
may help to shed some light on new types of molecules that are waiting
to be found in this region of the spectrum as well.

\section*{Acknowledgements}

This work is made possible by NASA Grant NNX17AH15G, NSF Grant
OIA-1757220, and the College of Liberal Arts at the University of
Mississippi. The computational facilities were provided by the
Mississippi Center for Supercomputer Research (MCSR).

\section*{Data Availability}

Supplementary data are available at MNRAS online.

\noindent
\textbf{Table S1.} CCSD(T)-F12b/cc-pVTZ-F12 equilibrium and first
vibrationally-excited rotational constants for \alo, \sin, and \sio\
(in MHz).

\noindent
\textbf{Table S2.} Fermi Resonances of \alo, \sio, and \sin.









\bsp	
\label{lastpage}
\end{document}